\documentclass{aa}      

\usepackage{epsfig,graphicx,url,twoopt}
\usepackage[varg]{txfonts}

\usepackage{natbib,twoopt}
\bibpunct{(}{)}{;}{a}{}{,} 

\begin{document} 

   \title{Disentangling 2:1 resonant radial velocity orbits from eccentric 
          ones and a case study for HD 27894
          \thanks{This research has made use of the Exoplanet Orbit 
          Database and the Exoplanet Data Explorer at 
          http://www.exoplanets.org .}}


   \author{Martin K\"urster\inst{1}
          \and
          Trifon Trifonov\inst{2}
          \and
          Sabine Reffert\inst{3}
          \and
          Nadiia M. Kostogryz\inst{4,5}
          \and
          Florian Rodler\inst{1,6}
          }

   \institute{Max-Planck-Institut f\"ur Astronomie,
              K\"onigstuhl 17, D-69117 Heidelberg, Germany\\
              \email{kuerster@mpia-hd.mpg.de flo@mpia.de}
         \and
              Department of Earth Sciences, The University of Hong Kong, 
              Pokfulam Road, Hong Kong, PR China\\
              \email{trifonov@hku.hk}
         \and
             Zentrum f\"ur Astronomie der Universt\"at Heidelberg, Landessternwarte,
             K\"onigstuhl 12, D-69117 Heidelberg, Germany\\
             \email{sreffert@lsw.uni-heidelberg.de}
         \and
              Kiepenheuer-Institut f\"ur Sonnenphysik,
              Sch\"oneckstr.~6, D-79104, Freiburg, Germany\\
              \email{kostogryz@kis.uni-freiburg.de}
         \and
             Main Astronomical Observatory of NAS of Ukraine, 27 
             Zabolotnoho str., Kyiv 03680, Ukraine\\
             \email{kosn@mao.kiev.ua}
         \and
             Harvard-Smithsonian Center for Astrophysics, 
             60 Garden St, Cambridge MA 02138, USA \\
             \email{frodler@cfa.harvard.edu}
             }

   \date{Received Month Day, Year; accepted Month Day, Year}

 
  \abstract
   {In radial velocity (RV) observations, a pair of extrasolar planets near a 2:1 orbital
    resonance can be misinterpreted as a single eccentric planet, if data are
    sparse and measurement precision insufficient to distinguish between these models.}
   {Using the Exoplanet Orbit Database (EOD), we determine the fraction of alleged 
    single-planet RV detected systems for which a 2:1 resonant pair of planets is also a 
    viable model and address the question of how the models can be disentangled.}
   {By simulation we quantified the mismatch arising from applying the wrong model.
    Model alternatives are illustrated using the supposed single-planet system HD~27894
    for which we also study the dynamical stability of near-2:1 resonant solutions.}
   {Using EOD values of the data scatter around the fitted single-planet Keplerians, we find 
    that for $74\% $ of the $254$ putative single-planet systems, a 2:1 
    resonant pair cannot be excluded as a viable model, since the error due to  
    the wrong model is smaller than the scatter.  For $187$ EOD stars  
    $\chi ^2$-probabilities can be used to reject the Keplerian models
    with a confidence of $95\% $ for $54\% $ of the stars and with $99.9\% $ for $39\% $
    of the stars.
    For HD~27894 a considerable fit improvement is obtained when adding a low-mass 
    planet near half the orbital period of the known Jovian planet.
    Dynamical analysis demonstrates that this system is stable when both planets are initially 
    placed on circular orbits.  For fully Keplerian orbits a stable system is only 
    obtained if the eccentricity of the inner planet is constrained to $<0.3$.}
   {A large part of the allegedly RV detected single-planet systems should be 
    scrutinized in order to determine the fraction of systems containing near-2:1
    resonant pairs of planets.  Knowing the abundance of such systems will allow us to revise
    the eccentricity distribution for extrasolar planets and provide direct constraints for 
    planetary system formation.}


   \maketitle

\titlerunning{2:1 eccentric orbits vs.~eccentric orbits}
\authorrunning{K\"urster et al.}
%

\section{Introduction}

With radial velocity (RV) measurements of the stellar reflex motion 
caused by orbiting companions a pair of planets in low-eccentricity
orbits near a 2:1 mean motion resonance (MMR) can be misinterpreted 
as a single planet with moderate eccentricity 
(\cite{2010ApJ...709..168A}; \cite{2013ApJS..208....2W}). 
This is, in particular, possible when the available data are sparse
and have large errors, when the stellar RV amplitude induced by the 
inner planet is smaller than that induced by the outer planet, 
and when the overall scatter of the data around either model is too 
large to make it possible to distinguish them.

A one-planet system might intuitively be considered a simpler model than 
a two-planet system and would therefore be favoured when applying Occam's 
razor to select the simplest hypothesis.  However, pairs of planets near a 
2:1 MMR are not rare and so warrant consideration.  
\citet{2011ApJS..197....8L} find that at least $16\% $ of the 
systems with more than one candidate for a transiting planet in the data from 
the Kepler satellite mission (e.g.~\cite{2010Sci...327..977B}) include a 
pair of planets with period ratios in the range 1.82:1 -- 2.18:1, i.e.~within
$9\% $ of the `pure' 2:1 ratio.  \citet{2014A&A...570L...7D}
find an excess of planets with a period ratio a few percentage points higher
than the 2:1 or 3:2 resonant value in data from the Q1-Q16 KOI catalogue 
(\cite{2013ApJS..204...24B}).  This effect is most pronounced for systems 
with periods of the inner planet $<5~\mathrm{d}$, for which no systems in the
exact resonance are actually found, it is still significant for inner planet
periods between $5~\mathrm{d}$ and $15~\mathrm{d}$, but strongly reduced
for longer periods, for which the exact resonance is
observed more frequently.  In their more recent analysis of Kepler 
data, \citet{2015MNRAS.448.1956S} find a significant tendency for pairs of
planets to have a period ratio close to 2.2:1.  In RV data, the true period 
ratio will be difficult to distinguish from the MMR when data are sampled 
sparsely and when they only cover a few planetary orbits.

Furthermore, it is interesting to 
note that the model of a two-planet system in precisely 2:1 resonant circular 
orbits (hereafter abbreviated as `2:1-RCO') has the same 
number of free parameters as the model of a single planet in a Keplerian 
eccentric orbit.  
In the case of the 2:1-RCO, the stellar RV varies as a function of time $t$ as

\begin{eqnarray}
RV(t) & = \gamma + \kappa_1\cos\left( 2\pi ~{t-t_\mathrm{max,1}\over P}\right)   
+\kappa_2\cos \left( 4\pi ~{t-t_\mathrm{max,1} \over P}-\Delta \phi \right) \\
      & = \gamma + \kappa_1\cos\left( 2\pi ~{t-t_{\mathrm{max},1}\over P}\right)
+\kappa_2\cos \left( 4\pi ~{t-t_{\mathrm{max},2}\over P}\right) ~.~~~~~~~
\end{eqnarray}

\begin{figure*}
\centering
\includegraphics{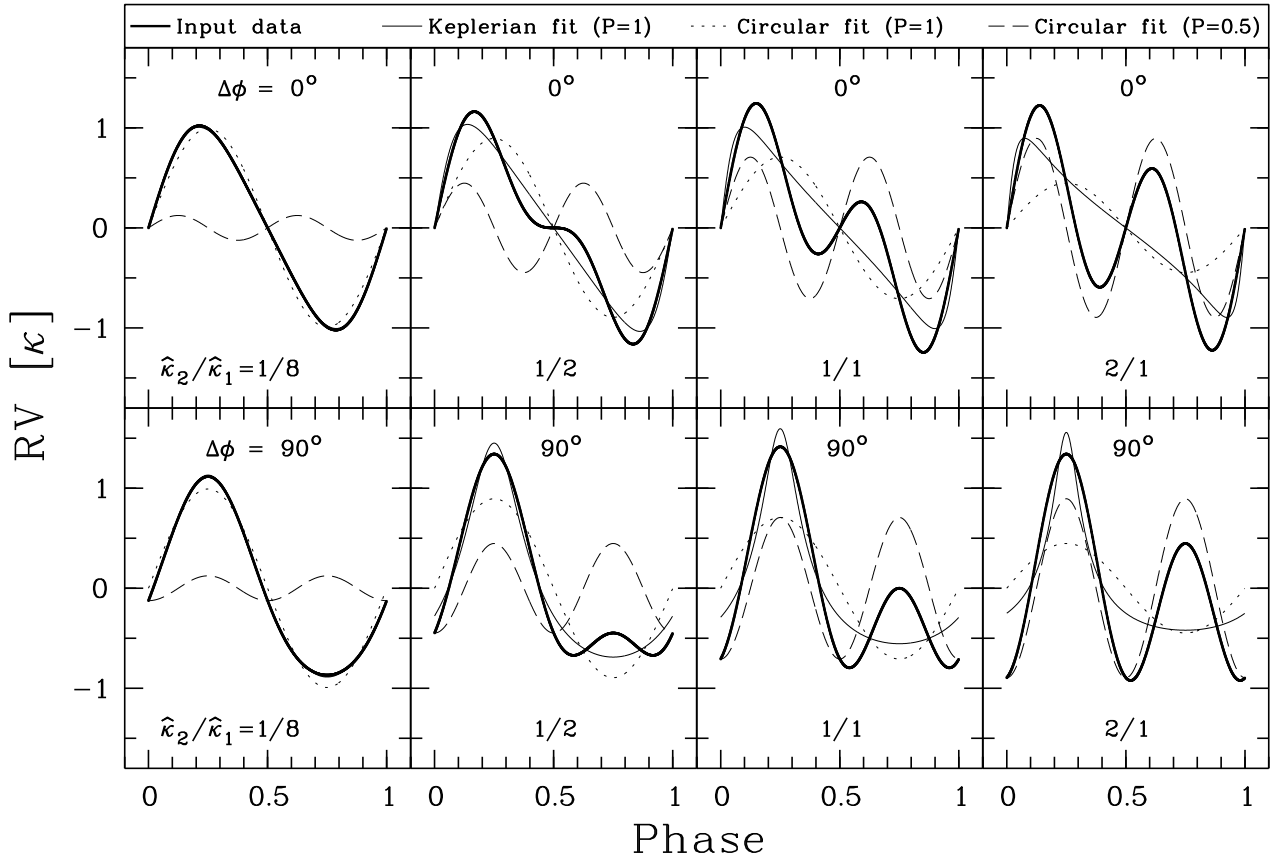}
\caption{Examples for single-component fits to the RV data of a 
pair of 2:1-RCO planets.  Thick solid lines
represent the input data which are shown for amplitude ratios
$\hat{\kappa}_2/\hat{\kappa}_1=1/8,~1/2,~1/1,$ and $2/1$ (from left to 
right) and values of the phase shift $\Delta \phi =0^\circ $ (upper row 
of panels) and $90^\circ $ (lower row of panels).  Thin solid lines depict
the best-fitting Keplerian orbits with $P=1$ (not discernible in the 
leftmost panels because of their very good match to the thick solid line).  
Dotted lines indicate best-fitting circular orbits (sine waves) with $P=1$.
Dashed lines stand for best-fitting circular orbits with $P=0.5$. 
Note that the best-fitting Keplerian orbits with $P=0.5$ are identical to 
the circular orbits with the same period, since the eccentricity 
vanishes for all fits.
Plots for phase shifts of $180^\circ $ and $270^\circ $ can be 
obtained by flipping the $0^\circ $ and the $90^\circ $ plots, 
respectively, around the horizontal axis and then shifting 
them by $0.5$ in phase.
} \label{FigCurves}%
 \end{figure*}

\noindent
Here $P$ is the period of the outer planet, $\kappa_1$ and $\kappa_2$ are 
the RV semi-amplitudes of the star induced by the outer and inner planet,
respectively, $\gamma $ is the systemic RV, and $\Delta \phi $ is the phase 
difference between the two sinusoids.  Finally, $t_\mathrm{max,1} $ and 
$t_\mathrm{max,2} $ are the times of the RV maximum that would occur if there 
were only the outer, respectively, the inner planet in the system.
\footnote{One may be inclined to attribute index `1' to the
inner planet and index `2' to the outer one.  We chose
the opposite sequence because in this paper we consider
`planet~1' as the one that has been shown to definitely exist, whereas the 
existence of potential `planet~2' has yet to be proven.  Also, in the 
cases we are interested in, the RV signal of the second planet (if any) can 
be regarded as a minor disturbance of the signal from the first.}

In the case of a single planet in an eccentric orbit the stellar reflex 
motion is given by

\begin{equation}
RV(t)=\gamma + K~{\sin E\sin \omega ~-\sqrt{1-e^2}\cos E\cos \omega 
\over 1-e\cos E}~, \\
\end{equation}

\noindent 
Here $E$ is called the `eccentric anomaly', which is related to the
`mean anomaly' $M$ via the Kepler equation

\begin{equation}
E-e\sin E=M=2\pi {(t-t_\mathrm{p})~\bmod P \over P}~.
\end{equation}

\noindent 
After introducing the `true anomaly' $\theta $ given by

\begin{equation}
\tan \theta = \sqrt{1-e^2}~{\sin E \over \cos(E-e)}~,
\end{equation}

\noindent 
the expression for $RV(t)$ can be written as

\begin{equation}
RV(t)=\gamma + K~[\cos (\theta +\omega ) + e\cos \omega ]~.
\end{equation}

The free parameters of this Keplerian model are the 
orbital period $P$, the stellar RV semi-amplitude $K$, the orbital eccentricity 
$e$, the longitude of periastron $\omega $, the time of 
periastron passage $t_\mathrm{p} $, and the systemic RV $\gamma $.

\begin{figure*}
\centering
\includegraphics{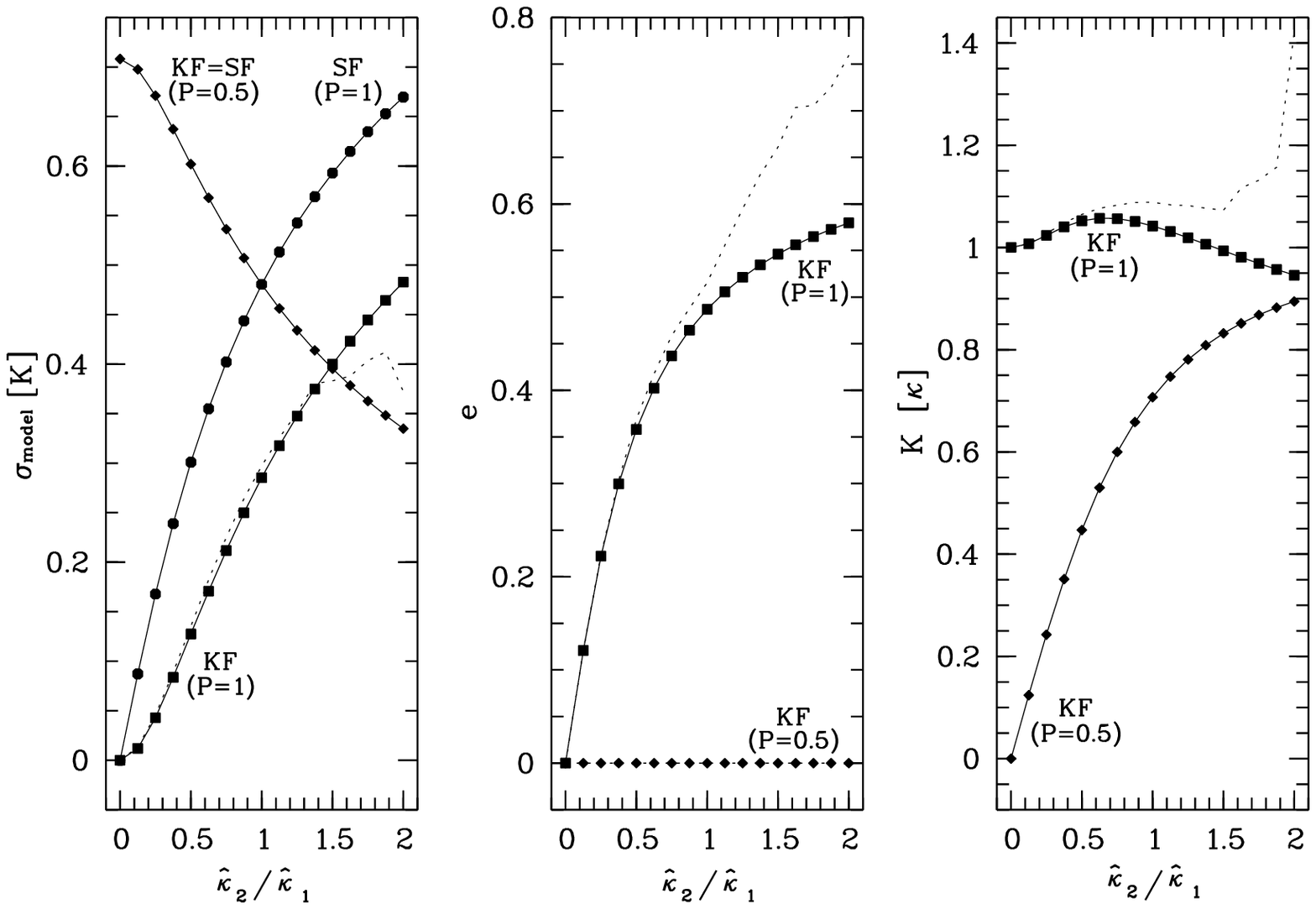}
\caption{Results from our simulations of fitting 2:1-RCO data with various
models.  Left panel: The rms deviation $\sigma _\mathrm{model}$ between the 
various types of fit and the 2:1-RCO input data as a function of the amplitude 
ratio $\hat{\kappa}_2/\hat{\kappa}_1$ of the input data.  Shown is 
$\sigma _\mathrm{model}$ for the single-component Keplerian fits (`KF') and for 
the sine fits (`SF') pertinent to the circular models.  Note that 
$\sigma _\mathrm{model}$ is expressed in units 
of the semi-amplitude $K$ of the Keplerian fit with Period $P=1$.  
~~---~~Middle panel: Variation of the eccentricity $e$ of the 
Keplerian fits with $\hat{\kappa}_2/\hat{\kappa}_1$.  
~~---~~Right panel: The semi-amplitude $K$ of the 
Keplerian fits as a function of $\hat{\kappa}_2/\hat{\kappa}_1$ and normalized 
to one for $\hat{\kappa}_2/\hat{\kappa}_1=0$.  In all panels fits with 
period $P=1$ are plotted with larger symbols than fits with $P=0.5$ (see the
labels at the curves).  Thin dotted lines in all three panels: 
As a comparison we have included the result from a
simulation for sparse data sampling based on the times at which the 
$20$~RV measurements for the host star of HD~27894 were taken by 
\citet{2005A&A...439..367M}. These thin dotted lines 
correspond to the median values for $\sigma _\mathrm{model}$, $e$, 
and $K$, respectively, obtained from this simulation (Keplerian 
fits with $P=1$ only). Note that for $\hat{\kappa}_2/\hat{\kappa}_1$
less than $\approx 0.5$ they are located near the line for 
KF $(P=1)$ for $e$ and $K$, whereas for $\sigma _\mathrm{model}$
the similarity holds up to $\hat{\kappa}_2/\hat{\kappa}_1\approx 1.4$.
}
\label{FigParams}%
 \end{figure*}

Both models have six free parameters.  It is a known difficulty with 
Occam's razor that the term `simplicity' is not easy to define.  On the 
one hand, the mathematics behind the Keplerian orbit is more complicated than 
that of circular orbits, since one has to solve the transcendent equation
for the eccentric anomaly $E$ (Eq.~4); also the description of the 
Keplerian orbit (Eqs.~3 and 6) involves more trigonometric functions 
than that of the 2:1-RCO pair of planets (Eqs.~1 and 2).
On the other hand, the number of assumptions leading to Eqs.~(1) or (2) is 
larger than for the single eccentric planet, since the eccentricities
of both planets of the pair must (for practical purposes, i.e.~within the 
measurement errors) be exactly zero and the period ratio must be exactly 2:1.  
If these assumptions are not made, and in reality they cannot be fulfilled 
exactly, the pair of planets has more model parameters than the single 
planet model, namely eleven vs.~six for the case that the system is well
described by two Keplerians.  In general, however, a realistic model for a 
pair of planets is even more complicated when the gravitational interaction 
between the two planets is considered.  Their mutual gravitational disturbance
leads to evolving orbital parameters 
(e.g.~\cite{2013ApJS..208....2W} and references therein; 
\cite{2014A&A...568A..64T}) which are unlikely to be found occupying the 
exact values required for the 2:1-RCO case at any time.

The discussion which model is simpler and therefore to be preferred has,
of course, a precursor in history important for our overall worldview, 
namely the question whether Kepler's elliptical orbits were to be preferred 
over Ptolemy's epicycles when it came to describing the observed positions 
of those of the solar system planets known at the times of these historical 
astronomers.  \footnote{The epicycles were an attempt to come up with a 
description that was completely based on (nested) circular movements, but 
capable of describing what in reality were Keplerian orbits.  The models we 
are considering here differ somewhat from that historical situation.  
Here we apply Keplerian orbits to single planets, whereas circular orbits 
are applied to pairs of planets in concentric 2:1 resonant motion.  In order 
to describe the behaviour of the solar-system planets well enough, two-component 
epicycles had to be `eccentric' with the main cycle centred
on a point different from Earth and the second cycle centred somewhere 
on the main cycle and moving along it.  Another difference is that we are 
applying our models to RV data which were not available yet to 
Ptolemy or Kepler who used astrometric measurements.}

Resonant orbits are found in the solar system as well as in extrasolar
planetary systems.  The most famous solar system examples for 2:1 resonant 
orbits are Jupiter's moons Ganymede, Europa, and Io which are in a configuration
called the Laplace resonance with orbital period ratios very close to 4:2:1.
Other examples for 2:1 resonances in the solar system are Saturn's moons Tethys 
and Mimas as well as Dione and Enceladus.  Also, the orbital period ratio of 
Neptune and Uranus deviates by less than $2\% $ from the 2:1 resonance.  
As an example for the possible destabilizing effect of resonances the Cassini 
division in Saturn's rings should be mentioned which has been cleared by a 
1:2 resonance with the moon Mimas.  Concerning the fact that resonances with 
massive bodies can stabilize or destabilize orbits we note that, on the one 
hand, the Trans-Neptunian Objects display various types of resonances with 
Neptune among which also the 2:1 ratio is found in several cases; on the 
other hand 1:2 resonances of Main-Belt asteriods with Jupiter are very rare.


\begin{table}
\caption{Fit mismatch for different models as a function of amplitude ratio.
The mean rms deviations $\sigma _\mathrm{model}$ of the model fits from the 
simulated input data are given in the 3rd -- 8th table column
for the amplitude ratios 
$\hat{\kappa}_2/\hat{\kappa}_1$ (1st column) used for Fig.~1.  
Means are taken over the 12 probed phases.
Each $\sigma _\mathrm{model}$ value is given in two different units, first in units 
of the RV semi-amplitude $K$ of the Keplerian fit with $P=1$ (3rd, 5th, and 7th 
column) and second in units of the amplitude $\kappa _1$ of the sine wave
with $P=1$ corresponding to the outer planet of the simulated  
2:1-RCO system (4th, 6th, and 8th column; see Eq.~7).  The 2nd column lists the 
values of the eccentricity pertinent to the Keplerian fit with $P=1$.
}
\label{table:1}
\centering
\begin{tabular}{c|c|c|c}
\hline
\hline
Ampl. & Keplerian fit & Sine fit       & Keplerian \\
ratio &               & (circular fit) & or Sine fit \\
input &               &                & \\
data  & $P=1$         & $P=1$          & $P=0.5$ \\
\hline
$\hat{\kappa}_2/\hat{\kappa}_1$ 
  & $e~~~~~~~~~~~~\sigma _\mathrm{model}~~~$ & $\sigma _\mathrm{model}$ 
                                           & $\sigma _\mathrm{model}$ \\
  & ~~~~~~~~~~~~~$[K]~~~~[\kappa _1]$~ & ~$[K]~~~~~[\kappa _1]$~~ 
                               & ~$[K]~~~[\kappa _1]$~~ \\
\hline
$1/8$ & $0.12$~~~~$0.012$~$0.012$    & $0.087$~~$0.089$ & $0.70$~~$0.71^{(b)}$ \\
$1/2$ & $0.36$~~~~$0.13$~~~$0.15~~$  & $0.30$~~~~$0.36~~$ & $0.60$~~$0.71^{(b)}$ \\
$1/1$ & $0.49$~~~~$0.29$~~~$0.42~~$  & $~0.48$~~~~$0.71^{(b)}$ & $0.48$~~$0.71^{(b)}$ \\
$2/1$ & $0.58$~~~~$0.48$~~~$1.0~~~~$ & $~0.67$~~~~$1.4^{(a)}~~$ & $0.34$~~$0.71^{(b)}$ \\
\hline
\end{tabular}
\begin{tabular}{llll}
\footnotesize 
Notes: & $^{(a)}$ \footnotesize $=\sqrt{2}$. & $^{(b)}$ \footnotesize $=1/\sqrt{2}$. 
       & \phantom{XXXXXXXXXXXXX111} \\
\hline
\end{tabular}
\end{table}


The Laplace resonance is also found for the components e, b, and c 
of the planetary system around the M4V dwarf GJ~876
(\cite{2010ApJ...719..890R}; \cite{2001ApJ...556..296M}) consisting of
two Jovian planets in a 2:1 resonance (the more massive one being in the outer 
orbit of the pair) plus a third low-mass planet in the 4:2:1 resonant orbit 
further out.  The system has a fourth, non-resonant Super-Earth-type planet in 
a very tight inner orbit around the host star.  Further examples for 2:1
resonant pairs of planets include the system around the G0V star HD~82943 with 
two Jovian planets (\cite{2004A&A...415..391M}) of similar minimum mass.  
Indications presented by \citet{2008MNRAS.385.2151B} for a third lower-mass 
planet in an outer orbit and in a 4:2:1 resonance with the inner two were 
refuted by \citet{2013ApJ...777..101T}.  A pair of two Jovian planets near 
the 2:1 resonance was also found in the system of the K0V star HD~128311, the 
outer one being more massive.  There is also evidence for a third inner 
non-resonant Saturn-type planet (\cite{2014ApJ...795...41M};
\cite{2005ApJ...632..638V}; \cite{2003ApJ...582..455B}).  Remarkable
in this system are the relatively high eccentricities of $0.303$ and $0.159$ 
for the inner and outer planet, respectively, of the resonant pair.  Around 
the K-giant $\eta $~Cet, \citet{2014A&A...568A..64T} found a system 
consisting of two Jovian planets whose period ratio is within 
$\approx 9\% $ of the 2:1 resonance.  Another special system is KOI-730 whose 
four supposed Super-Earth planets seem to have period ratios of 3:4:6:8,
which thus include two interlaced occurences of the 2:1 resonance 
(\cite{2011ApJS..197....8L}).

In the present paper we investigate the possibility that some fraction 
of the extrasolar planets in eccentric orbits found by the RV technique
are actually pairs of planets in orbits that are close to circular and near a 
2:1 resonance.  By simulating and comparing the respective orbits as a 
function of their model parameters (Sects.~2 and~3) we explore 
the circumstances under which the two models are indistiguishable.
This enables us to identify those of the known eccentric planets in the 
literature for which suitable follow-up observations could determine 
which model is correct.  To illustrate this we study one such 
example (Sect.~4).  In the following discussion we provide recommendations 
for the strategy for such new observations that may uncover a so 
far unknown additional planet near the 2:1 resonant orbit (Sect.~5).
Finally, we summarize our main conclusions (Sect.~6).


\section{Simulations}


We have simulated RV data for pairs of 2:1-RCO planets for the purpose of 
fitting them with single-component Keplerian models as well as with 
single-component circular models (i.e.~sine waves).
For this we used the fitting routines provided by the software for the 
Generalized Lomb Scargle (GLS) periodogram 
\citep{2009A&A...496..577Z}. 
The simulated data $RV_i$ are composed of two components
$RV_{1,i}$ and $RV_{2,i}$ and calculated from

\begin{equation}
RV_i=RV_{1,i}+RV_{2,i}=\kappa_1\cos(2\pi t _i)+\kappa_2\cos(4\pi t _i 
-\Delta \phi)~.
\end{equation}

\noindent 
A comparison with Eq.~(1) shows that we have chosen the period of the outer
planet as $P=1$ (hence the period of the inner planet is 0.5) and set the
uninteresting parameters to $t_\mathrm{max,1}=0$, and $\gamma =0$.  


\subsection{Dense sampling}

As a numerical approximation to continuous RV curves,
we created densely sampled data sets containing 1000 points calculated 
at equidistant sampling times $t_i$ and filling the interval $[0,P)$, 
i.e. the sampling times are given by the values
$0.000,~0.001,~...~,0.999$ and provide an even distribution of the 
sampled phases.  We produced these data with the 
following 17 values of the ratio of semi-amplitudes:
$\kappa_2/\kappa_1 = i/8$ for $i=0,...,16$, 
i.e~$\kappa_2/\kappa_1 = 0$, $1/8$, $1/4$, $3/8$, ... , $2$.
The semi-amplitudes in Eq.~(7) were always normalized according to

\begin{equation}
\hat{\kappa}_1=\kappa_1 /\kappa ~~~~~\mathrm{and}~~~~~\hat{\kappa}_2=\kappa_2 /\kappa
~~~~~\mathrm{with}~~~~~\kappa=\sqrt{\kappa_1^2 + \kappa_2^2}~,
\end{equation}

\noindent 
where $\hat{\kappa}_1$ and $\hat{\kappa}_2$ denote normalized values, whereas 
$\kappa_1$ and $\kappa_2$ stand for non-normalized values.

The simulated data were subsequently fitted with four different
single-component models:

\begin{enumerate}
\item sine fit with a period of $P=1$,
\item sine fit with a period of $P=0.5$,
\item Keplerian fit with a period of $P=1$,
\item Keplerian fit with a period of $P=0.5$.
\end{enumerate}

For each value of the semi-amplitude ratio of the inner over the outer planet 
$\kappa_2/\kappa_1=\hat{\kappa}_2/\hat{\kappa}_1$ we produced $12$ data sets 
with the $12$ different, equally spaced phase shift values 
$\Delta \phi = 0^\circ,~30^\circ,~...~,330^\circ.$  A natural 
exception to this is the data set with $\hat{\kappa}_2/\hat{\kappa}_1=0$ 
for which the phase shift is not defined.

In order to compare the fits, their rms deviations $\sigma _\mathrm{model}$ 
from the simulated 2:1-RCO data were employed to indicate the quality of the 
fit.  This is justified since 
our simulations are based on idealized data without errors, and we compare
the optimum fits of either type without any uncertainties in the fit 
parameters.  However, whenever we draw any conclusions from real data 
(Sects. 4 and 5) we instead compare the probability of chi-square 
$p(\chi ^2)$ of the fits.  We express rms values for both the Keplerian fits 
and the sine fits in units of the semi-amplitude $K$ obtained for
the Keplerian fit with $P=1$.  


\begin{figure}
\resizebox{\hsize}{!}{\includegraphics{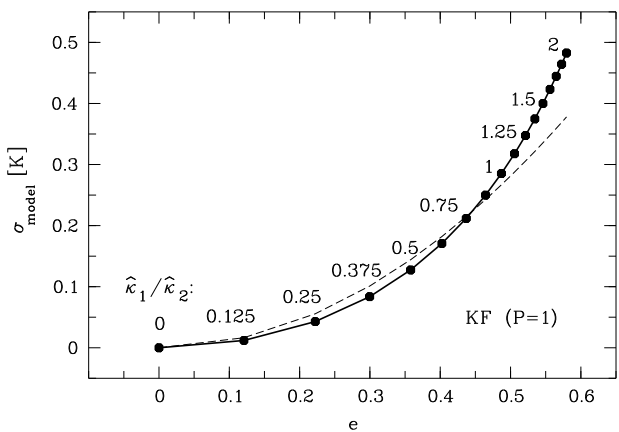}}
\caption{Rms deviation 
$\sigma _\mathrm{model}$ of the Keplerian fit with $P=1$ from the 2:1 resonant
input model (solid line with points and labels).  This deviation is 
expressed in units of the RV semi-amplitude $K$ and plotted as a function of 
the eccentricity $e$ determined by the fit.  The labels indicate the 
amplitude ratio $\hat{\kappa}_2/\hat{\kappa}_1$ of the input sinusoids.  
Thin dashed line: amplitude of the first harmonic in the Fourier expansion 
of the Kepler equation.
}
\label{FigEccentricity}%
 \end{figure}


\subsection{Sparse sampling}

In order to compare the results from our idealized densely sampled data sets 
with those that are obtained from real observations, we created rather sparsely 
sampled data sets consisting of only 20 data points 
sampled at the temporal pattern of the observations of the planet host star 
HD~27894 (\cite{2005A&A...439..367M}).  These observations are quite unusual
due to the small number of data points that led to the discovery of this planet.

With this sparse sampling grid we simulated again 2:1-RCO RV data with the 
same $17$ values for the amplitude ratio $\kappa_2 / \kappa_1$ and the same 
$12$ phase shift values $\Delta \phi $ as for the densely sampled data set 
described in Sect.~2.1.  We then fitted these data, but restricted ourselves 
for the purpose of this comparison to only the Keplerian model with period 
$P=1$.  Due to the larger number of free parameters compared with the
sinusoidal fits the strongest discrepancies are expected for the Keplerian fits.


\section{Results}

\subsection{Single-planet fits to 2:1-RCO planets}

Fig.~1 shows examples of single-planet model fits to the densely 
sampled simulated input data with different amplitude ratios 
$\hat{\kappa}_2/\hat{\kappa}_1$ and two selected values of
the phase shift $\Delta \phi = 0^\circ$ and $90^\circ $, while Tab.~1 shows 
the mean rms deviation $\sigma _\mathrm{model}$ of these model fits from the 
data where the mean has been taken over all 12 phase shift values
(see Sect.~3.2 on the relatively small variation of the individual values
of $\sigma _\mathrm{model}$ as a function of the phase).
Actually, as can be seen from Fig.~2 (left panel), the single Keplerian always
yields the best fit, i.e.~the smallest $\sigma _\mathrm{model}$, for 
$\hat{\kappa}_2/\hat{\kappa}_1<1.5$, while for 
$\hat{\kappa}_2/\hat{\kappa}_1\ge 1.5$ the fit with $P=0.5$ is better.


\begin{figure}
\resizebox{\hsize}{!}{\includegraphics{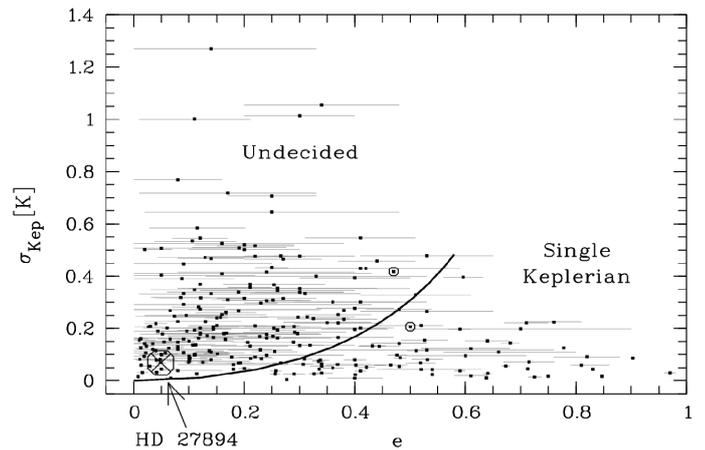}}
\caption{Rms scatter $\sigma _\mathrm{Kep}$ of the RV solutions 
around single-planet Keplerian fits for our sample of 254 stars for which the 
EOD provides these values (dots).
The scatter values are plotted as a function of the formal
eccentricity of the Keplerian fit and in units of its RV semi-amplitude.    
$1\sigma$--errors of the eccentricity are represented by grey horizontal 
bars.  In two instances no errors are available; 
the pertinent data points are marked by small circles around them.
The large circle and cross marks the case of HD~27894 that is studied
in Sect.~4.3.  The solid curve is identical to the solid line in Fig.~3 
representing the deviation $\sigma _\mathrm{model}$
between the single-planet Keplerian model and the model for the 
2:1 resonant pair of planets.  Scatter values above the curve are
larger than the expected deviation arising from applying the erroneous
model (single Keplerian instead of 2:1 resonant pair) so that both models
are possible, whereas scatter values below the curve are too small to 
assume an erroneous model and therefore favour the single Keplerian.
}
\label{FigEccStars}%
 \end{figure}


Specifically, for an amplitude ratio $\hat{\kappa}_2/\hat{\kappa}_1=1/8$ 
(Fig.~1, left panels), the data (thick solid line) and the Keplerian fit with 
the period of the outer planet (thin solid line) become indistiguishable in 
our plot, whereas the single sinusoidal fit with the same period (dotted line)
can be seen to systematically deviate from the data, and the fit with the 
period of the inner planet (dashed line) is completely inadequate.
For an amplitude ratio of $1/2$ (Fig.~1, second column of panels), the 
single Keplerian with the period of the outer planet deviates systematically
from the data, but could still be a reasonable match to real data, if 
signal-to-noise ratios are moderate.  For this amplitude ratio, the 
sinusoidal fit with the period of the outer planet deviates much more 
from the data, and the fit with the period of the inner planet deviates 
even more (but less than for the amplitude ratio of $1/8$).
For an amplitude ratio of $1/1$ (Fig.~1, third column of panels),
all single-planet fits are inadequate, but the Keplerian fit with the 
period of the outer planet is still the formally best fit.  For this 
amplitude ratio, the sine fits with $P=1$ and $P=0.5$ are of 
equally poor quality since they have the same rms deviation (see Tab.~1).
If the amplitude ratio of the inner vs.~the outer planet is reversed then for
$\hat{\kappa}_2/\hat{\kappa}_1\ge 1.5/1$ the fit with the period of the inner 
planet is the best 
one, followed by the Keplerian fit with the period of the outer planet, 
and then the sine fit with the period of the outer planet (see Fig.~1, 
right panels which are for $\hat{\kappa}_2/\hat{\kappa}_1=2/1$).


\subsection{Limited dependence on phase shift}  

For the densely sampled data sets, we find that the interesting fit parameters 
RV semi-amplitude $K$ and eccentricity $e$, and also the rms of the fit vary 
as a function of phase shift $\Delta \phi $ only by very small 
amounts.\footnote{This is in agreement with the Fourier expansion of the 
Kepler equation given by \citet{2010ApJ...709..168A} in their Eqs.~(1) and 
(2) which show that up to first order the Keplerian orbit is identical to the 
2:1 pair of sinusiods (with $\omega $ assuming the role of $\Delta \phi $).}
For the extreme amplitude ratio $\hat{\kappa}_2/\hat{\kappa}_1=2$, these 
variations are $0.55\% $, $1.5\% $, and $8.0\% $ peak-to-valley, respectively, 
for the rms, for $K$, and for $e$.  For smaller amplitude ratios, these values 
are even smaller.
Therefore, in our analysis, for every given amplitude ratio, we are using
the mean values for the rms and for $e$ and $K$ obtained from the fits to the
12 data sets with different phase shifts (see Sect.~3.1).  For the sparsely 
sampled data sets, 
we use the median values (instead of the mean values) of $e$ and $K$ for all 
further comparisons, since they display somewhat stronger variations and can 
have strong outliers so that the median is the more representative value.


\subsection{Zero eccentricity in P = 0.5 fits}

We also find that in the case of the Keplerian models with period $P=0.5$ all 
fits to the densely sampled data sets yield an eccentricity $e=0$ and an
amplitude and rms equal to that of the sine fits with the same period value.
That is why we will henceforth no longer distinguish between Keplerian fits 
and sine fits for $P=0.5$.  As noted above, we have not made any fits with 
$P=0.5$ to the sparsely sampled data sets.  In this case, the Keplerian 
fit with $P=0.5$ will normally show differences 
from the circular fit, since the sparse sampling does no longer guarantee
the symmetric distribution of the radial velocities which is naturally present 
for a densely sampled sine wave.
However in this paper, we will largely concentrate on cases of small
deviations from the single sinusoidal model, such as low-eccentricity 
Keplerians or small-amplitude second circular planets with
half the period of the (circular) first one.  Consequently, we are exploring 
the regime where $P=0.5$ fits alone are never a good match anyway.

\subsection{Dependence of rms, e, and K on amplitude ratio for dense sampling}

This behaviour can also be deduced from Fig.~2 which shows the
dependence of the rms of the different types of fits on the amplitude
ratio of the input data (left panel).  Fig.~2 also shows the variation
of the eccentricity $e$ (middle panel) and the semi-amplitude $K$ of the 
Keplerian fit (right panel) with this ratio.

\subsection{Example for dependence of rms, e, and K on amplitude ratio for sparse sampling}

In the three panels of Fig.~2, thin dotted lines show the results from the
data sets simulated using the sparse sampling example outlined in Sect.~2.2.  
With this example, we illustrate that conclusions concerning the 
goodness-of-fit, the determined eccentricity, and the RV semi-amplitude drawn 
from our idealized densely sampled data are likely to hold also for 
realistic data sets as long as the RV amplitude ratio 
$\hat{\kappa}_2/\hat{\kappa}_1$ of the 2:1-RCO planets
is not too large.  In particular, our sparse data example and our dense data 
sets produce similar $\sigma _\mathrm{model}$ values for amplitude ratios 
$\hat{\kappa}_2/\hat{\kappa}_1$ up to $\approx1.4$, and similar $e$ and $K$ 
values for $\hat{\kappa}_2/\hat{\kappa}_1$ up to $\approx 0.5$ (see Fig.~2).

\subsection{Dependence of rms on e -- a first tool to search for erroneous
models}

Fig.~3 shows the rms deviation of the Keplerian fit with $P=1$ from the 
simulated 2:1-RCO data 
as a function of the pertinent value for the eccentricity obtained 
by the same fit.  This is based on the densely sampled data.  
We will use this relation as a tool to identify those of the known 
RV determined planetary systems for which single objects on eccentric 
orbits have been fitted, but a pair of 2:1 resonant planets is also a 
possible model.

Looking, in observed data (with not too sparse sampling and moderate
amplitude ratios; see Sect.~3.5), at the scatter $\sigma _\mathrm{Kep}$ of the 
RV measurements around the fitted Keplerian model (expressed in units of the 
RV semi-amplitude $K$ of the fit), Fig.~3 allows us to determine the 
expected (typical, average) contribution $\sigma _\mathrm{model}$ 
to the scatter arising from the wrong model fit.  This means that all
systems that we can identify in the literature which have a planet in a
putative eccentric orbit as well as a scatter of the data around the 
eccentric model larger than the one given by the solid curve in Fig.~3 
are potential candidates for a 2:1 resonant pair of planets.  

In principle, the instrumental noise $\sigma _\mathrm{inst}$ and the 
RV jitter $\sigma _\mathrm{star}$ arising from stellar effects such as 
activity, convective motions, and pulsations, commonly named `stellar jitter,'
should also be added in quadrature to the scatter due to the wrong model 
fit.  Systems with an observed RV scatter $\sigma _\mathrm{Kep}$ exceeding 
this combined scatter would then be candidates for a follow-up study in search 
for a 1:2 resonant second planet, i.e.~sytems with 
$\sigma _\mathrm{Kep} ^2 \ge \sigma _\mathrm{model}^2 
+\sigma _\mathrm{inst}^2 +\sigma _\mathrm{star}^2$.
In practice, however, this can be done only for those systems for which
good estimates of the instrumental noise and of the stellar noise are 
available.  Estimates of the instrumental noise can often be obtained
from a joint analysis of the data of the ensemble of stars observed in a
survey-type programme, e.g.~by looking for the most constant stars, but 
even for those substantial effort is required to disentangle the 
instrumental noise from stellar jitter.  Short-term jitter due to stellar
p-mode oscillations can be estimated using the scaling relation from 
\citet{1995A&A...293...87K}.  
There is also quite some literature on estimates of the 
effects of stellar activity (e.g.~\cite{2011A&A...528A...4B}; 
\cite{1997ApJ...485..319S}; \cite{1998ApJ...498L.153S}; 
\cite{2000ApJ...534L.105S}), but because of the nature of this activity 
(spot migration, activity cycles, etc.) they cannot be very precise.


\subsection{Probability of chi-square -- the tool of choice for real data}

As already alluded to in Sect.~2.1, it should be noted that the 
scatter alone is not a statistically 
sound criterion, since in general, it is possible to find signals that are 
smaller than the scatter/noise, if one has a sufficiently large number of
measurements.  Rather than looking at the rms, the appropriate approach 
would be to look at the probability of $\chi ^2$ of the fitted models,
and in fact our main data base for planetary orbital parameters (the EOD; 
see Sect.~4) provides values of $\chi ^2$ (along with the also required 
values for the degrees of freedom, DoF) for a good fraction of the interesting 
entries, but only so for the Keplerian fit,
not for an alternative fit with a 2:1 resonant pair of sinusoids. 


\subsection{First harmonic of the Fourier expansion of the Kepler equation}

\citet{2010ApJ...709..168A} base much of their study of 2:1 resonant 
vs.~Keplerian models on the Fourier expansion of the Kepler equation.
While we have adopted a different approach in our simulations, a comparison 
is in order at this point.
It is given by the dashed line in Fig.~3 which represents the amplitude
of the first harmonic of the Fourier expansion of the Kepler equation
(from \cite{2010ApJ...709..168A} and \cite{2005A&A...439..663L}). 
The similarity between the two curves in Fig.~3 demonstrates that the ability 
to distinguish between the Keplerian model and the model with the 2:1 resonant 
pair of circular orbits is largely equivalent to the ability to detect 
this first harmonic.


\begin{figure}
\resizebox{\hsize}{!}{\includegraphics{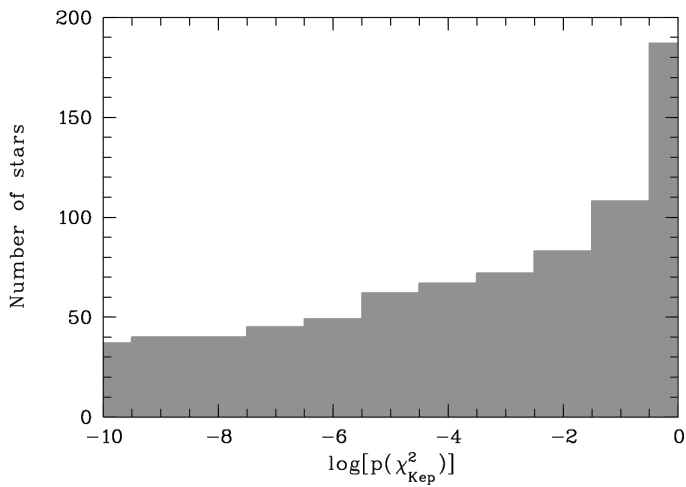}}
\caption{Cumulative histogram of $\log[p(\chi _\mathrm{Kep} ^2)]$ of the 
Keplerian fits for those $187$ stars in our sample for which both the reduced 
chi-square $\chi ^2_\nu$ of the fit and the number of measurements is 
provided by the EOD.  For a substantial fraction of the stars, the 
chi-square probability is too small to accept the single
Keplerian model, e.g.~$39\% $ of the stars have 
$\log[p(\chi _\mathrm{Kep} ^2)]<-3$ (see also Tab.~2).
}
\label{FigCumul}%
 \end{figure}

\subsection{The model-related scatter in real data}

We also note that the so determined model-related contributions 
$\sigma _\mathrm{model}$ to the scatter are exact only for very densely sampled 
RV curves such as those we have used for our simulations (Sect.~2.1).  
For much more sparsely sampled real data, the model-related contributions 
to the scatter can deviate somewhat in the individual case from the here 
determined values.
Typically, data sets with fewer measurements can be fitted `better'
\footnote{Better, that is, in terms of the rms scatter, but not if fit 
criteria such as the probability of chi-square, $p(\chi^2)$ can be adopted.} 
by any multi-parameter model, so that the model-related contribution to the 
rms scatter is likely to be somewhat smaller than what we have determined for 
our simulated data sets with $1000$ equidistant data points.   
Nevertheless, we will conservatively use these larger values to guide us in 
our search for candidate systems (i.e.~systems with even larger scatter).


\section{Follow-up candidates from the literature}

\subsection{Sample definition}

In order to come up with a compilation of candidate systems, we used the 
Exoplanet Orbit Database (EOD) described by \citet{2011PASP..123..412W}
and available on-line at http://www.exoplanets.org.
From the list of detected exoplanets, we have selected two versions of 
a sample of stars based on the following criteria.

\begin{enumerate}
\item Stars with an RV orbital solution.
\item Stars for which no planetary transits have been observed.
\item $e>0$.
\item Listed number of components equal to one (supposedly single planets).
\item $K$ must be given.
\item a) Values for the rms scatter of the data around the 
      \phantom{a)~~}Keplerian fit $\sigma _\mathrm{Kep}$ are available.\\
      b) Values for the probability of chi-square 
      $p(\chi _\mathrm{Kep}^2)$ can be
      \phantom{b)~~}obtained from the data base as it provides both the re- 
      \phantom{b)~~}duced $\chi ^2$ and the number of observations.
\end{enumerate}

As of January 13th, 2015, we find that $254$ of the stars contained in the EOD 
fulfill criteria $1-5$ plus 6a; these will make up our version 1 sample.
$187$ of the stars fulfill criteria $1-5$ plus 6b and will constitute our
version 2 sample.  $185$ stars fulfill criteria $1-5$ plus 6a and 6b.
Concerning criterion 6b, we note that the EOD contains entries for the (reduced) 
chi-square $\chi _\mathrm{Kep}^2$ and the number of measurements $N$ from which 
we determine $p(\chi _\mathrm{Kep}^2)$ using standard routines 
(e.g.~\cite{1969drea.book.....B}).\footnote{For the calculation of 
$p(\chi _\mathrm{Kep}^2)$, one must know the number of degrees of 
freedom (DoF) which we take as $N-6$, because of the six free parameters of the
Keplerian model.  In a few cases, however, the listed $\chi _\nu ^2$ values 
belong to models that also include a linear or curved trend.  Then the DoF 
will be smaller by one or two, and $p(\chi ^2)$ will be 
a little larger than the values we have used.  However, this is a small 
effect which we ignore here.  It matters only when $N$ is small.}

Note that in order to constrain the sample, we discarded known transiting
systems (criterion 2), since for those chances are higher that a second planet 
further inside in the system has already been detected through its own transits.  
We also restricted 
ourselves to supposedly single-planet systems (criterion 4), because finding an 
additional low-amplitude planet in a 2:1 resonance with one of the planets 
in the system will be more difficult when more than one planet has already 
been found to contribute to the observed stellar RV signal.

\begin{table}
\caption{Suitability of the single Keplerian model alone, i.e.~assuming that 
no other sources of variability exist.  Second row: 
number of stars for which the chi-square probability of the
single Keplerian model is less than the values given in the first row.
Third row: fraction of the $187$ stars in our version 2 sample.
Last row: confidence $C=1-p(\chi ^2_\mathrm{Kep})$ with which the single 
Keplerian model can be rejected.
}
\label{table:2}
\centering
\begin{tabular}{lcccc}
\hline
\hline
\noalign{\smallskip}
$p(\chi ^2_\mathrm{Kep})$ & $<0.05$ & $<0.01$ & $<0.005$  & $<0.001$  \\
\noalign{\smallskip}
\hline
\noalign{\smallskip}
No.~of stars            & $101$    & $83$     & $79$      & $72$      \\
Fraction                & $54\% $  & $44\% $  & $42\%   $ & $39\% $   \\
Confidence $C$          & $>95\% $ & $>99\% $ & $>99.5\% $ & $>99.9\% $ \\
\noalign{\smallskip}
\hline
\end{tabular}
\end{table}

\begin{figure*}
\includegraphics[height=6.45cm]{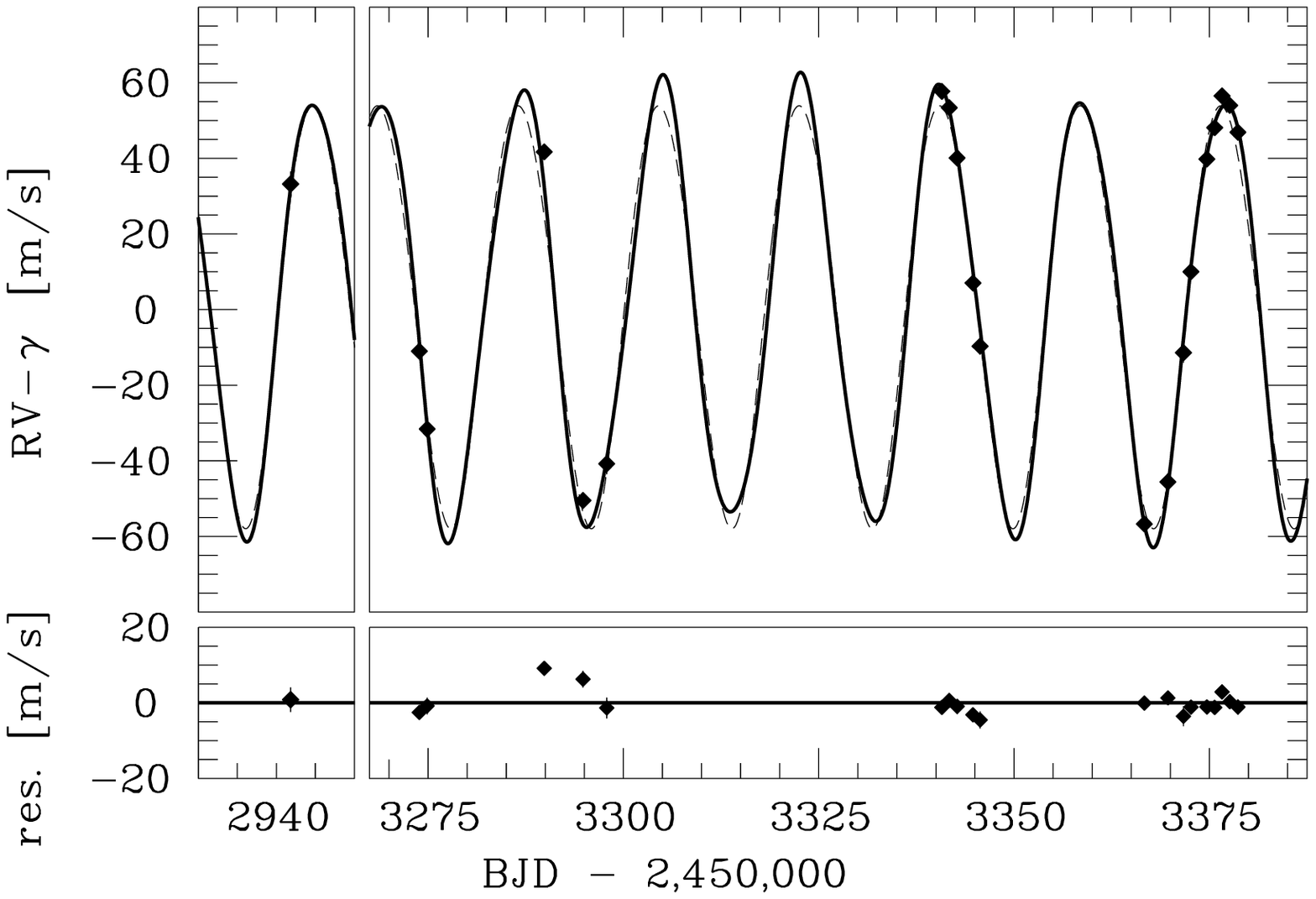}
\includegraphics[height=6.45cm]{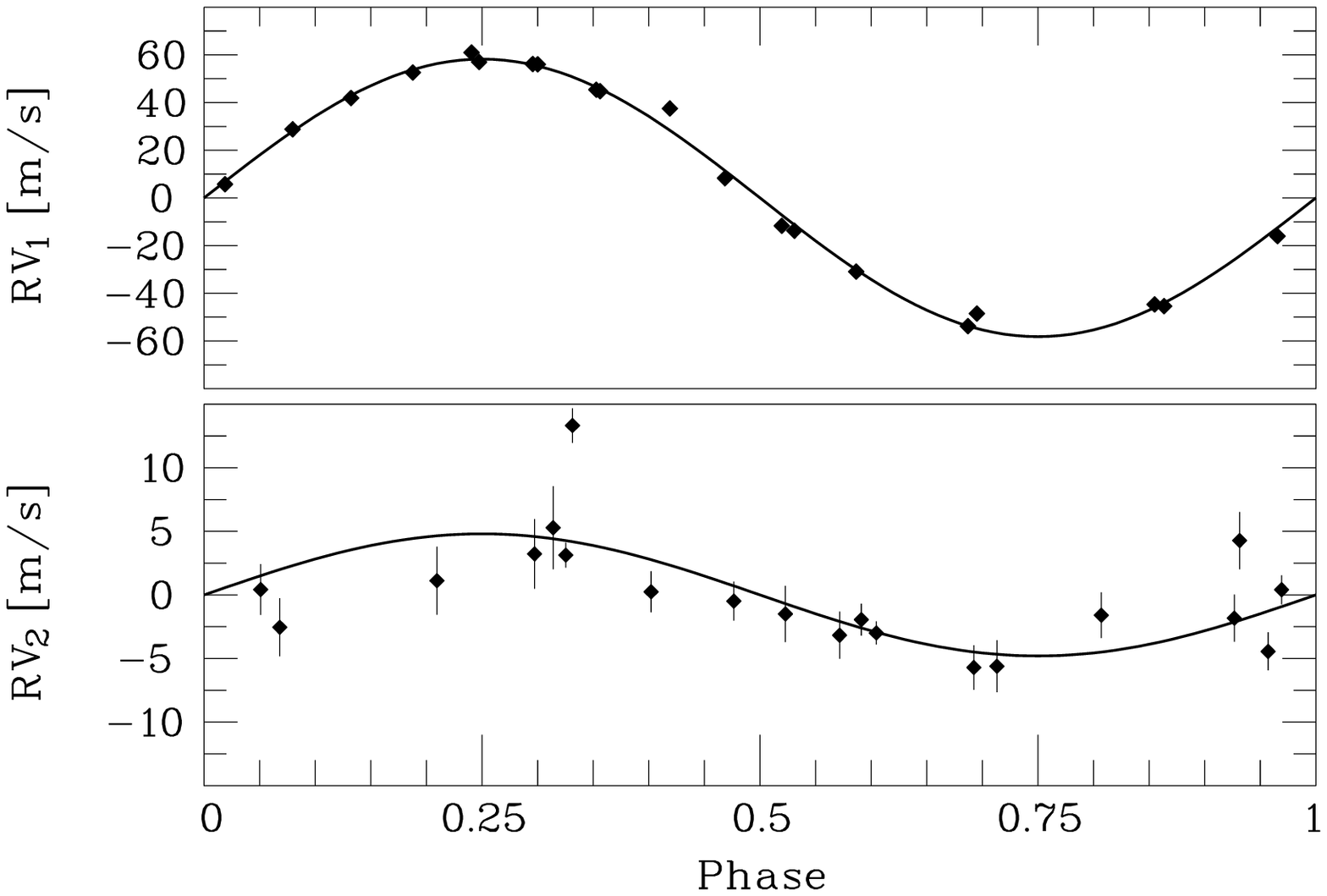}
\caption{Fit to the HD~27894 RVs with a model consisting of two planets in circular
orbits.  Note that in this case the exact 2:1 resonance is not forced, but the two
period values are both free parameters of the fit.  ~~---~~Left: Time series together with
the best-fitting two-planet circular model (top, thick solid line) and residuals (bottom).
For display purposes the x-axis has been broken between
BJD - 2,452,950 and BJD - 2,453,267.5 where there is a large data gap.
The thin dashed line in the upper panel represents the best-fitting single Keplerian model
for comparison.  ~~---~~Right: RV data phase folded with the period of the outer planet 
after subtraction of the signal of the inner planet (top) and phase folded with the 
period of the inner planet after subtraction of the signal of the outer planet (bottom).
}
\label{FigHD27894model}%
\end{figure*}

\subsection{Resulting candidate sample}

Fig.~4 shows the location in the rms-vs.-eccentricity diagram of our version 1
sample stars in comparison to the curve from Fig.~3, i.e.~the curve that 
depicts the expected contribution to the rms if a 2:1 resonant system with 
two circular orbits is fitted with a single-star Keplerian model.  We find that 
$188$ of the $254$ stars in the sample are located above the curve, 
i.e.~$74\% $ of the stars are possible candidates for hosting a system of 
two planets in a 2:1 resonance when the criterion of exceeding the 
model-related scatter for 2:1-RCO systems is applied (see Sect.~3.6).

In order to apply also the probability of $\chi ^2$ of the single Keplerian fits
as a criterion for identifying candidates (Sect.~3.7), we display in Fig.~5 
(on a logarithmic abscissa) the cumulative histogram of 
$p(\chi _\mathrm{Kep}^2)$ 
for our version 2 sample stars which demonstrates that a large fraction of the 
stars has a very low probability that the applied model alone is correct.
This is also shown by Tab.~2 that lists how many stars have 
$p(\chi _\mathrm{Kep}^2)<0.05,~<0.01,~<0.005$, and $<0.001$, respectively,
and which fraction of the total of $187$ stars in
our version 2 sample this corresponds to.  It also provides the confidence
for the rejection of the single Keplerian model for these stars given by 
$1-p(\chi _\mathrm{Kep}^2$).  The hypothesis that the Keplerian model alone
can explain the observed variability can be rejected for $54\% $ of our 
version 2 sample stars at the $95\% $ confidence level and for
$39\% $ at the $99.9\% $ confidence level.

\subsection{A case study: HD~27894}

From our (version 1) sample of candidate 2:1-RCO systems (systems above the
curve in Fig.~4), we selected the 
20 very sparsely sampled RV measurements of HD~27894 (Tab.~A.2 in 
\cite{2005A&A...439..367M}) in order to provide an example for a 
detailed study of a supposedly single-planet system which might hide a 
second planet in an inner orbit near a 2:1 MMR with the first one.  
(We have already employed the observing times of this star in Sect.~2.2.)
HD~27894 is a low-activity K2V star with a mass of $0.8~\mathrm{M}_\odot $.
Further characteristics of the star are described in 
\citet{2005A&A...439..367M}.

HD~27894 was selected because the moderate eccentricity of the published
Keplerian fit as well as the considerable scatter of the data around it made 
this system a perfect candidate for an ambiguous, undecided case well suited 
for an illustrating example.  This system earns much of this ambiguity from 
its very sparse data sampling.  

We base our analysis of the HD~27894 system on various models.
First we apply a single Keplerian fit just like \citet{2005A&A...439..367M}
and obtain fit parameters in agreement with those found by these authors
(except for a $2~\mathrm{m~s}^{-1}$ offset in the systemic RV $\gamma $;
see Tab.~3).  As discussed in Sect.~1 this model has six free parameters 
(Eqs.~3 and 4).  We find a low-eccentricity fit 
with a reduced chi-square of $\chi ^2_\mathrm{\nu } = 8.706$ (DoF=$14$), hence
$p(\chi ^2) = 2.69\times 10^{-19}$ implying that the fitted model alone 
cannot at all describe the data if the measurement errors are reliable.
At this point, of course, a multi-planet model is not necessarily called for
yet, since stellar RV jitter (due to pulsations, convection, and activity) could 
also be responsible for the excess variability; see the pertinent discussion
in \citet{2005A&A...439..367M} on the fact that the adopted average RV
error of $1.8~\mathrm{m~s}^{-1}$ does not take stellar variability into account. 
The single Keplerian fit has an rms of $4.21~\mathrm{m~s}^{-1}$ and yields a minimum 
companion mass of $m\sin i=0.645~\mathrm{M}_\mathrm{Jup}$ and an orbital 
period of $P=17.99~\mathrm{d}$ (Tab.~3).  

Next we consider two-planet fits.  We start off with the exact 2:1-RCO case
which yields values for the orbital period and planetary mass for the outer 
planet very close to those found for the single-Keplerian model (Tab.~3)
and only a small improvement of chi-square, $\chi ^2_\nu = 8.581$, 
$p(\chi ^2)=5.92\times 10^{-19}$ (DoF $=14$) and rms ($4.17~\mathrm{m~s}^{-1}$).
In this model, the inner planet has $m\sin i =8.4~\mathrm{M}_\mathrm{Earth}$.
The similarity of the 2:1-RCO fit with the
single Keplerian demonstrates the ambiguity of the model
selection in this case of sparsely sampled RV data.

The situation changes somewhat when we leave the two period 
values free but retain circular orbits (eccentricities $e_1=e_2=0$, hence
$\omega _1$ and $\omega _2$ are undefined).  Our new model now has
seven free parameters, one more than Eq.~(1) or (2) since here we are not
forcing the 2:1 ratio for the periods.  The resulting fit corresponds to a 
two-planet system with $P_1 = 17.9919~\mathrm{d}$, 
$m_1\sin i=0.648~\mathrm{M}_\mathrm{Jup}$, $P_2 = 8.276~\mathrm{d}$, and
$m_2\sin i=0.044~\mathrm{M}_\mathrm{Jup}=14~\mathrm{M}_\mathrm{Earth}$ 
(see Tab.~3 and Fig.~6).  The two-planet fit has a substantially smaller
$\chi ^2_\mathrm{\nu } = 5.444$ (DoF=$13$), but still a very small 
$p(\chi ^2)$ of $2.37\times 10^{-19}$, probably since stellar jitter is 
dominating the noise behaviour but is not considered in the model.  
The rms of the fit is $3.20~\mathrm{m~s}^{-1}$.
Applying an F-test to the $\chi ^2$ ratio of the fits, we find a relatively 
moderate confidence of $83\% $ that this new model is an improvement over either
the single Keplerian or the 2:1-RCO models (same confidence value in both cases).
If the F-test is applied to the rms's of the fits, a similar value of 
$82\% $ is obtained in both cases.

We then study the orbital stability of the possible two-planet system 
following an approach similar to the one outlined in 
\citet{2014A&A...568A..64T}.  We perform N-body simulations employing the
symplectic integrator {\em SyMBA} \citep{1998AJ....116.2067D}
which is particulary suited to treat close encounters between the 
planets.  
The fitting code is described in detail in \citet{2013ApJ...777..101T}.

For these simulations we assumed co-planar 
orbits seen edge-on for both planets, i.e.~the orbital inclinations are 
$i_1=i_2=90^\circ $.  The initial orbital elements are obtained in the 
Jacobi frame matched to the representation of hierarchical multi-planet systems 
(see \cite{2003ApJ...592.1201L}).  
Our choice of co-planar edge-on orbits can be motivated as 
follows. First, if nothing is known about the orbital inclination and random 
orientation is assumed, $90^\circ $ is the most probable value.  Second, for 
this inclination value the planet mass is equal to its minimum mass determined 
from the RV data.  While higher mass planets will have a stronger gravitational
interaction and can potentially disturb each other more, edge-on orbits 
are the relevant ones for the question whether there are stable orbits at all.
Third, while co-planarity maximizes the interaction between the planets, 
co-planar orbits are nevertheless thought to be the 
predominant configuration for planetary systems originating from a 
circumstellar disk.  This leads us to assume them here.


\begin{table*}
\caption{Orbital parameters for the HD~27894 system for the various applied models.  $N_\mathrm{par}$ is the number of model parameters and 
$\mathrm{DoF}=N_\mathrm{data}-N_\mathrm{par}$ the number of degrees of freedom (with $N_\mathrm{data}$ the number of data points).  All model 
parameters from $P_1$ to $\gamma $ are as defined for Eqs. (1-4).  Framed values were held fixed for the fits.  $m_1\sin i$ and $m_2\sin i$ 
are the derived minimum masses of the outer and inner planet, respectively, $a_1$ and $a_2$ the pertinent semi-major axes of the orbits.}
\label{table:3}
\centering
\begin{tabular}{l|c|c|ccc}
\hline
\hline
 & \multicolumn{2}{|c|}{1-planet models}& \multicolumn{3}{|c}{2-planet models} \\
 & Moutou et al. & This work & \multicolumn{3}{|c}{This work} \\
 & (2005)        &           & 2:1-RCO     & 2-pl. circular & \\
 & single         & single    & $e_1=e_2=0$ & $e_1=e_2=0$ & Double \\
 & Keplerian      & Keplerian & $P_1=2P_2$  & $P_1\not = 2P_2$ & Keplerian \\
\hline
\hline
$N_\mathrm{par}$  &  6 &  6 &  6 &  7 & 11 \\
\hline
$P_1$~~~~~~~~~~$[\mathrm{d}]$ 		& $17.991\pm 0.007$ 	&$17.9904\pm 0.0077$ & $17.9915\pm 0.0071$ & $17.9919\pm 0.0070$ & $17.99\pm 0.35$ \\
$P_2$~~~~~~~~~~$[\mathrm{d}]$ 		& --- 			& ---  		     & ~\fbox{$P_1/2$}   & $~8.276~~\pm 0.035~$ & $~~8.27\pm 0.99$ \\
$t_\mathrm{p,1}$~~~~~~~~~$[\mathrm{BJD}-2450000]$ & $2933.7\pm 0.5^{(a)}$ & $2933.64\pm 0.55$ & ~~---  & ~~---            & $2928.2\pm 1.6$ \\
$t_\mathrm{p,2}$~~~~~~~~~$[\mathrm{BJD}-2450000]$ & ---               	& --- 		     & ~~--- & ~~---           	& $2933.8\pm 5.9$ \\
$t_\mathrm{max,1}$~~~~~~$[\mathrm{BJD}-2450000]$    & ---                   & ---                & $2926.98\pm 0.16$   & $2927.03\pm 0.16$  & --- \\
$t_\mathrm{max,2}$~~~~~~$[\mathrm{BJD}-2450000]$    & ---                   & ---                & $2939.40\pm 0.25$   & $2941.60\pm 1.75$  & --- \\
$e_1$            				& $0.049\pm 0.008$ 	& $0.0486\pm 0.0086$ & ~~\fbox{$0$} & ~~\fbox{$0$} & $0.057\pm 0.021$ \\
$e_2$            				& ---             	& ---  		     & ~~\fbox{$0$} & ~~\fbox{$0$} & $0.532~\pm 0.076~$ \\
$\omega _1$~~~~~~~~~~~~~$[^\circ $] 			& $132.9\pm 9.7~~$ 	& $132\pm 11$        & ~~--- 		 & ~~--- & $20.1\pm 24.6$ \\
$\omega _2$~~~~~~~~~~~~~$[^\circ $] 			& ---             	& ---  		     & ~~--- 		 & ~~--- & $30.8\pm ~~8.0$ \\
$K$ or $\kappa_1^{(b)}$~~~~$[\mathrm{m~s}^{-1}]$ & $58.1\pm 0.5$ 	& $58.13\pm 0.50$    & $58.02\pm 0.50$ 	 & $58.28\pm 0.49$ 	& $57.57\pm 0.60$ \\
$\kappa_2$ or $K_2^{(b)}$~~$[\mathrm{m~s}^{-1}]$ & ---           	& ---  		     & $~~3.00\pm 0.51$  & $~~5.17\pm 0.62$ 	& $~~8.24\pm 1.33$ \\
$\gamma $~~~~~~~~~~~~~~~$[\mathrm{km~s}^{-1}]$ 	& $82.9023$             & $82.90033$         & $82.90225$  	 & $82.90231$    & $82.90179$ \\
                                              & $\pm 0.0003$          & $\pm 0.00057$      & $\pm 0.00036$ 	 & $\pm 0.00039$ & $\pm 0.00053$ \\
\hline
$m_1\sin i^{(c)}$~~$[\mathrm{M}_\mathrm{Jup}]$	& $0.62$ 		& $0.645$            & $0.645$	 & $0.648$ 	& $~~0.639$ \\
$m_2\sin i^{(c)}$~~$[\mathrm{M}_\mathrm{Jup}]$	& ---   		& ---    	     & $0.026$   & $0.044$  	& $~~0.060$ \\
$m_2\sin i^{(c)}$~~$[\mathrm{M}_\mathrm{Earth}]$ & ---   		& ---    	     & $8.4$     & $14$         & $19$ \\
$a_1^{(c)}$~~~~~~~~~~~~$[\mathrm{AU}]$ 		& $0.122$        	& $0.125$            & $0.125$   & $0.125$	& $0.125$ \\
$a_2^{(c)}$~~~~~~~~~~~~$[\mathrm{AU}]$ 		& ---             	& ---		     & $0.079$   & $0.074$	& $0.074$ \\
\hline
rms~~~~~~~~~~~$[\mathrm{m~s}^{-1}]$ 	& $4.0$ 		& $4.21$ 	     & $4.17$ 	 & $3.20$ 	& $2.42$ \\
$\chi ^2_\nu $  		& not given 		& $8.706$  	     & $8.581$   & $5.444$ 	& $2.522$ \\
DoF             		& $14$ 			& $14$ 		     & $14$	 & $13$ 	& $9$ \\
$p(\chi ^2_\nu)$ 		& not given 		& $2.69\times 10^{-19}$ & $5.92\times 10^{-19}$ & $2.37\times 10^{-19}$ & $6.91\times 10^{-3}$ \\
\hline
\hline
Stability & n.a. & n.a. & Yes & Yes & No$^{(d)}$ \\
\hline
\end{tabular}
\begin{tabular}{ll}
\footnotesize 
Notes: & $^{(a)}$ \footnotesize Moutou et al.~(2005) quoted a value of $3275.5\pm 0.5$, which is later by $19$ revolutions of the planet.\\
       & $^{(b)}$ \footnotesize As in the text, $K$ stands for the semi-amplitude of the single Keplerian, wheras $\kappa_1$ and $\kappa_2$ stand for the 
                               semi-amplitudes of the\\
       & \footnotesize ~~~~~circular orbits of the outer and inner planet, respectively; $K_2$ is meant to denote the semi-amplitude of the inner planet in the\\
       & \footnotesize ~~~~~double Keplerian fit (last column).\\
       & $^{(c)}$ \footnotesize All determinations of $m\sin i$ and $a$ made in this work use a stellar mass of $0.8~\mathrm{M}_\odot $.
                               Contrary to their claim of having used the\\
       & \footnotesize ~~~~~same value, Moutou et al.~(2005) must have adopted $0.75~\mathrm{M}_\odot $ in order to arrive at their somewhat
                       smaller $m_1\sin i$ and $a_1$.\\
       & $^{(d)}$ \footnotesize Stable only if $e_2$ is constrained to $<0.3$ (for the case $i_1=i_2=90^\circ $).\\
\hline
\end{tabular}
\end{table*}


\begin{figure}
\resizebox{\hsize}{!}{\includegraphics{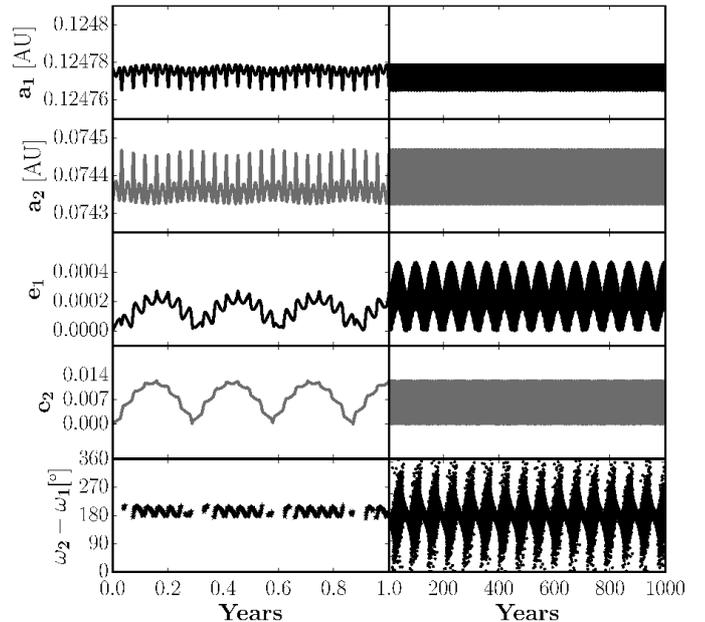}}
\caption{Orbital evolution of the best-fit system of two planets that started off on 
circular orbits but with unconstrained periods.  ~~---~~Left: 1-year excerpt from the 
simulation.  ~~---~~Right: excerpt for the following $999~\mathrm{yr}$.  Both in the left and 
right part of the figure,
the individual panels show from top to bottom the short-term variation of the semi-major
axes $a_1$ and $a_2$, of the eccentricities $e_1$ and $e_2$, and of the difference of the 
longitudes of periastron $\omega _2 - \omega _1$.  The last quantity is not plotted when
one of the eccenticities are close to zero ($e_1$ or $e_2<0.0001$) since then the 
longitude of periastron is undefined (hence the gaps in the lower left panel).
The variation pattern of these parameters shows various periodicites of which the
following can be most easily seen here:
$15~\mathrm{d}$ and $104~\mathrm{d}$ (all five parameters, left part of figure), 
and $65~\mathrm{yr}$ ($e_1$ and $\omega _2 - \omega _1$; right part of figure).
}
\label{FigHD27894_OrbitalEvolution}%
\end{figure}

We find the best fit with initially two circular orbits and free period values
to be stable over the complete simulation time span of $10000~\mathrm{yr}$ 
corresponding to over $2.3\times 10^5$ and $4.6\times 10^6$ orbits of the outer 
and inner planet, respectively.  A $1~\mathrm{yr}$ and a (contiguous) 
$999~\mathrm{yr}$ excerpt from the simulation are shown in Fig.~7 and reveal 
very regular oscillations (on several time scales) of the semi-major axes, 
the eccentricities, and the difference of the longitudes of periastron 
of the two planets.  The amplitudes of these oscillations are small.  
The semi-major axes of the outer and inner planet vary by about 
$10^{-4}$ and $2\times 10^{-4}$,
repectively.  Only very small eccentricity values 
are reached ($0\le e_1\le 0.00045$ and $0\le e_2\le 0.013$).
With a period ratio of $P_1/P_2 = 2.17$, this system is not exactly in 
2:1 MMR but still in the range of period ratios within $\pm 9\% $ of the 
MMR found to be frequent by \citet{2011ApJS..197....8L} in Kepler data 
(see Sect.~1 above).  It is particularly noteworthy that this period ratio
is very close to the value of 2.2 where a significant excess of planet 
pairs was found by \citet{2015MNRAS.448.1956S} also in Kepler data
(Sect.~1).

In order to check whether a system starting off at period ratios near 2:1 
(including the exact ratio) can also be stable, we examine a high density 
($50\times 50$ fits) $\chi ^2$ grid of two-planet circular fits with 
different initial periods.  We have tested each fit for long-term stability.
The result is shown in Fig.~8 which shows a modified chi-square 
defined to be equal to DoF at its minimum by 
$\chi ^2_\mathrm{mod}=\mathrm{DoF}\cdot \chi ^2/\chi ^2_\mathrm{min}$.  
Note that all models on this map are stable, 
and so is the one starting at the exact 2:1 period ratio which
is also consistent with the optimum fit (at $\chi ^2_\mathrm{min}$) 
at better than the $3\sigma $ level; more precisely, it 
lies within the combined $3\sigma $ error regime for the interesting 
parameters $P_1$ and $P_2$ around the optimum fit 
(cf.~\cite{1976ApJ...208..177L}; also \cite{1976ApJ...210..642A}).

Finally, we fit the eccentric two-planet model to the data.  First we 
apply the double Keplerian model, but we are unable to constrain the 
eccentricity of the inner planet (perhaps due to the sparse data sample).  
The best-fitting double Keplerian has $\chi ^2_\nu =2.522$ and 
$p(\chi ^2)$ of $6.91\times 10^{-3}$ (DoF $=9$; see Tab.~3 for the fit 
parameters).  When comparing it with the 6-parameter single Keplerian and 
2:1-RCO models, an F-test applied to the $\chi ^2$ ratios shows that in both
cases there is a confidence of $99\% $ that the five additional parameters 
of the double Keplerian provide a significant improvement of the fit.  
When the F-test is applied to the ratio of the squared rms's,
this confidence is $95\% $ in both cases.  However, 
the best-fitting double Keplerian yields a very large value for the
eccentricity of the inner planet of $e_2=0.51$.  Not only does this value
have an enormeous uncertainty of $\pm 0.47$, it also leads to a highly 
unstable planetary configuration.  In order to exclude such 
unstable models by imposing stability constraints, we apply a self-consistent 
two-planet N-body model 
(\cite{2001ApJ...551L.109L}; \cite{2013ApJ...777..101T}).
We find that, if we constrain the eccentricity of the inner planet to 
$e_2<0.3$, we arrive at a stable orbit again.

Comparing the models we have studied in this subsection we certainly find
some preference for a two-planet model for the HD~27894 system even though
this is not fully conclusive yet.  This is mainly due to the sparseness of 
the available data.
Since the formal best fit with two Keplerians drives the eccentricity of the 
potential inner planet to values so high that the system becomes unstable,
models with moderate eccentricities should be preferred.  This includes the 
model with two circular orbits.  HD~27894 is clearly an example for a system 
where more observations hold the potential to uncover a second planet.

\begin{figure}
\resizebox{\hsize}{!}{\includegraphics{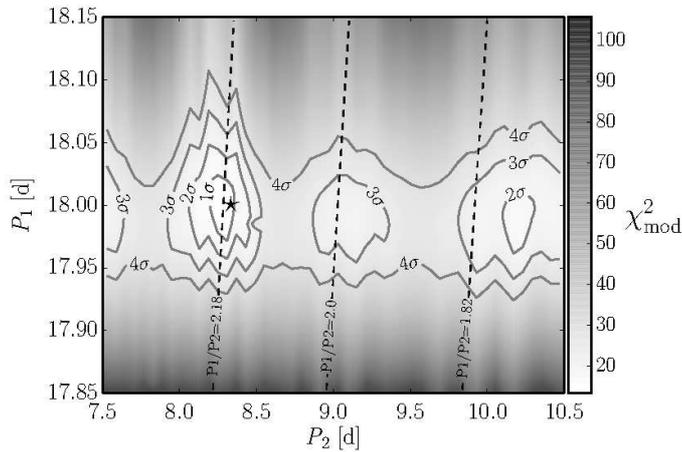}}
\caption{Map of modified chi-square, 
$\chi ^2_\mathrm{mod}=\mathrm{DoF}\cdot \chi ^2/\chi ^2_\mathrm{min}$, 
as a function of the two orbital periods of the two-circular-planets
model for HD~27894.  All models on this map are stable.  Contours are 
combined confidence levels with respect to the best fit, 
i.e.~the $1,~2,~3$ and $4~\sigma $ regimes in which both period 
values are found together with a probability of $68.27\% $, $95.45\% $, 
$99.73\% $, and $99.99\% $, respectively (while the remainder of the orbital 
parameters is optimized, i.e.~treated as uninteresting).  The best fit with 
$\chi ^2_\mathrm{mod}=\mathrm{DoF}=13$ is denoted by an asterisk.
It is within the $\pm 9\% $ region from the exact 2:1 period ratio. 
The left and right slanted dashed lines delimit this region whereas the
middle one marks the exact ratio.  
Exact 2:1 solutions are found within the $3~\sigma $ contour (middle).  
The $1,~2,~3$ and $4~\sigma $ contours correspond to 
$\chi ^2_\mathrm{mod}=15.30,~19.18,~24.83$, and $32.33$, respectively.
}
\label{FigHD27894_7par_stability}%
\end{figure}


\section{Discussion}

\subsection{Observational strategy for the follow-up}

In this paper we have illustrated how RV data with uncertainties can lead observers
to apply the wrong model for the description of the data of planet hosting 
stars.  In particular, we have concentrated on the topic of the erroneous 
application of a single Keplerian one-planet model instead of a 2:1 resonant 
two-planet model.  Sparse sampling as well as insufficient measurement precision 
can make it difficult to distinguish between the two models.  Sparse sampling 
may just not cover enough of those orbital phases where the two models differ 
most, and measurement errors may be too large for attempts to resolve the 
difference between the models.

Naturally, improvement will come from more observations of the candidate stars,
and even more so if also the measurement precision can be improved.
Observational methods other than RV measurements, such as astrometry or 
photometric transits can contribute to the selection of the best model for the 
studied planetary system.  \citet{2014ApJ...795...41M} provide an example
for employing constraints from astrometry and photometry to 
distinguish between or exclude certain models.
Mostly though, the follow-up will consist of securing more RV measurements.
These can be optimized based on the following recipe:

\begin{itemize}
   \item Fill phase gaps in the data.
   \item Preferentially observe those phases of the RV curve where the 
         differences between the models are strongest.
\end{itemize}

On average, doubling the number of measurements increases the ability to resolve
the difference between the models by a factor of $\sqrt{2}$.  With higher instrumental
precision this ability is even greater provided that the main limitation is not
due to intrinsic effects of the star such as activity, convective motions, and
pulsations.

To illustrate the best choice of phases by way of example, take the second panel 
from the left in
the upper row of panels in Fig.~1.  If one compares the thick solid line, i.e.~the
2:1-RCO two-planet model, with the thin solid line, i.e.~the Keplerian one-planet
model, the largest differences occur at phases near $0.18$, $0.40$, $0.60$, and $0.82$ 
whereas intermediate phases have smaller discrepancies.  Given the usually limited
observing time, one should concentrate on those phases of the largest differences.

\subsection{Biased eccentricity distribution and consequences for planetary system formantion}

Determining the fraction of two-planet systems near the 2:1 resonance among the
systems previously classified as eccentric single-planet ones will allow us to
find the degree to which the observed eccentricity distribution is biased.
It may turn out that solar-system-like architectures with low-eccentricity 
planets are more frequent than previously thought.
If a significant bias can be established, this will certainly have implications
for theories of the formation of planetary systems.
We note in this context also that, according to several authors, resonances 
observed in planetary systems provide clues to planetary migration processes 
and the architecture of those systems (e.g.~\cite{2014ApJ...795L..11P}; 
\cite{2014A&A...570L...7D}; \cite{2006MNRAS.365.1160B}).  

\subsection{Outer planets as perturbers}

One related topic is worth mentioning here, even though we have not studied
it in this paper. \citet{2009ApJ...702..716R} investigated the effect that 
undetected second planets in wide (outer) orbits have on the determined 
eccentricity value of supposed single planets in sparse RV data.  
These authors find that a massive outer planet can lead to published 
eccentricity values for the detected inner planet that are too large.
In this way they have a similar effect as the 2:1 resonant inner planet
on which we have concentrated in this study.


\section{Conclusions}

   \begin{enumerate}
      \item We have studied the case that in RV measurements a two-planet system with
            2:1 resonant circular orbits (2:1-RCO system) is erroneously 
            interpreted as a single-planet system and fitted with a single eccentric
            Keplerian.
      \item By way of simulation we have determined how, on average, the discrepancy 
            between the true and the false models depends on the eccentricity of 
            the fitted Keplerian.
      \item From the excess scatter of the data around the fitted model, we find that
            in $74\% $ of $254$ stars for which a single eccentric planet was announced 
            a 2:1-RCO system is a possible alternative model that cannot be excluded.
      \item Based on the probability of $\chi ^2$ we, find that for $54\% $ of $187$ stars 
            the Keplerian fit can be rejected as the sufficient model at the $95\% $ 
            confidence level and for $39\% $ of $187$ stars it can be rejected at the
            $99.9\% $ confidence level.
      \item Therefore, a large number of single-planet systems announced in the literature
            should be scrutinized with follow-up observations in order to provide the
            data quality needed to distinguish between the two models in question.
      \item Conceivably, it may then be further established that a substantial fraction of 
            the supposed
            single eccentric orbits are in fact near-circular systems of two planets near
            the 2:1 resonance.  This finding would imply a bias in the determined
            eccentricity distribution making the quasi-circular orbits prevailing in the
            solar system more frequent than presently thought.  This would have direct
            implications on planet formation theory.
      \item Employing the example of HD~27894, we illustrate the model alternatives finding
            that this star may have a Uranus or Neptune-mass planet in an inner orbit near
            the 2:1 resonance with its known Jovian planet.  Intriguingly, 
            our two-planet models for this star yield a period ratio close to 2.2, a value 
            where an excess of planets pairs is found in data from the Kepler mission.
      \item For this system, we also perform stability simulations showing that there
            are stable orbits at or near the 2:1 MMR when we assume initially circular 
            orbits for both planets.
      \item If fully Keplerian orbits are allowed for both the known and the hypothesized
            planet, the HD~27894 system is only stable when the eccentricity of the inner
            planet is constrained to $<0.3$.
      \item We suggest that follow-up observations of candidate systems should naturally 
            concentrate on those phases of the orbit of the known planet with the strongest 
            discrepancies between the two-planet and the single-planet (Keplerian) models.
   \end{enumerate}


\begin{acknowledgements}
We thank the anonymous referee for useful suggestions.
NMK ackowledges support by the European Research Council Advanced Grant 
HotMol (ERC-2011-AdG 291659).
FR acknowledges financial support from the Alexander von Humboldt foundation.
\end{acknowledgements}





\end{document}